# Symmetry-assisted protection and compensation of hidden spin polarization in centrosymmetric systems

Yingjie Zhang, Pengfei Liu, Hongyi Sun, Hu Xu and Qihang Liu*

*Shenzhen Institute for Quantum Science and Technology and Department of Physics, Southern University of Science and Technology, Shenzhen 518055, China*

*Email: liuqh@sustc.edu.cn

***Abstract***

It is recently noted that in certain centrosymmetric compounds, spin-orbit interaction couples each local sector that lacks inversion symmetry and thus leads to visible spin polarization effects in the real space, dubbed as "hidden spin polarization (HSP)". However, observable spin polarization of a given local sector suffers interference by its inversion partner, impeding material realization and potential applications of HSP. Starting from a single-orbital tight-binding model, we propose a nontrivial way to protect strong sector-projected spin texture through minimizing the interaction between inversion partners by nonsymmorphic symmetry. The HSP effect is generally compensated by inversion partners near the Γ point but immune from interaction around the boundary of the Brillouin zone. We further summarize 17 layer groups that support such symmetry-assisted HSP and identify by first-principle calculations hundreds of quasi-2D materials from the existing databases, among which a group of rare-earth compounds LnOI (Ln = Pr, Nd, Ho, Tm and Lu) serves as great candidates showing strong Rashba- and Dresselhaus-type HSP. Our finding also provides an ideal platform to control HSP for emergent physical effects, such as opening a hybridization gap by tensile strain to realize the time-reversal-invariant topological superconductivity.

**I. Introduction**

In nonmagnetic materials, spin polarization, measured by the difference between spin-up and spin-down electrons for a given direction $S = (I_\uparrow - I_\downarrow)/(I_\uparrow + I_\downarrow)$, was the exclusive territory of materials with strong spin-orbit coupling (SOC) and broken inversion symmetry [1]. Nevertheless, it was recently shown that certain centrosymmetric crystals formed by inversion-asymmetric sectors also manifest Rashba- or Dresselhaus-type spin textures when projecting the Bloch wavefunctions onto these local sectors [2]. Having been confirmed in a number of centrosymmetric bulk materials experimentally [3-6], such a form of "hidden spin polarization" (HSP) considerably broadens the range of materials for potential spintronic applications and brings exotic physical insights into the existing fields, such as spin field effect transistor [7], topological insulator [8] and topological superconductivity [9], etc. Very recently, HSP is reported in traditional cuprates and thus poses the open question of how the high-temperature superconductivity correlates with such nontrivial spin pattern [10]. Furthermore, the concept of HSP has triggered a broader field of "hidden polarization", where various physical effects were recognized to be determined by the local symmetry breaking of a system, albeit with a higher global symmetry that seemingly prohibit the effect to happen. Examples include orbital polarization [11], optical activity [12], circular polarization [13,14], and Berry curvature [15], etc.

Based on the simple symmetry analysis on the local symmetry of atomic sites or sectors [2], apparently most centrosymmetric materials would manifest HSP, especially for quasi-2D layered materials with finite thickness. However, materials with strong HSP that hold the realistic potential for applications indeed form a much smaller subgroup. For example, although silicon in diamond structure is a nominal Dresselhaus-type HSP system because each Si atom has an inversion-asymmetric $T_d$ site point group, the strong interaction between the two Si sublattice (through Si-Si bond) significantly compensates the HSP localized on each Si atom. Therefore, after completing "proof of existence" for HSP, the next important question is *which specific physical features within centrosymmetric crystals are actually controlling the magnitude of such HSP effect*, i.e., "where to look" [16]. Besides strong SOC that favors heavier elements, the design principle of HSP also requires minimizing the interaction between sectors. One obvious but trivial way is to separate the two sectors as far as possible. This is analogous to the thick slab of a topological insulator with inversion symmetry, in which the top and bottom surfaces manifest energy-degenerate Dirac cones and spin polarizations with opposite chirality [17]. Such type of HSP is basically two individual sets of spin polarization in quite different place, which is difficult to manipulate and integrate in one system.

Here, we explore a nontrivial, symmetry-assisted approach to minimize the interaction between the inversion-asymmetric sectors and thus protect the HSP in each sector. Our tight-binding model reveals that when the two sectors connect each other by nonsymmorphic crystalline symmetry, not only the time-reversal invariant momumtum (TRIM) at the Brillouin zone (BZ) boundary holds four-fold degeneracy, but also the sector interaction vanishes along the BZ boundary and is strongly suppressed along other directions from TRIM points, in consistent with the recent first-principle calculation on a specific material $BaNiS_2$ [18]. As a result, a spin map constructed throughout the full BZ shows that nearly perfect HSP survives even the two sectors are close to each other in a quasi-2D lattice. We further perform symmetry analysis and identify 17 quasi-2D layer groups (out of 80) in total that support such symmetry protected HSP. Among the realistic materials in the existing quasi-2D material databases [19,20] with the selected layer groups, we perform first-principle calculations and choose a rare-earth family LnOI (Ln = Pr, Nd, Ho, Tm and Lu) as representative candidates showing strong Rashba and Dresselhaus HSP, which is in excellent agreement with our tight-binding model. Our finding offers a general principle for extensively exploring strong HSP materials, and also provides an ideal platform to control HSP for emergent physics and applications. One example is to tune by tensile strain the hybridization gap of such a HSP system for realizing time-reversal-invariant topological superconductivity [9,21,22].

## II. Results

### A. Tight-binding model

We begin with a simple model of a nonsymmorphic 2D lattice to illustrate how HSP

in one sector survives under the existence of its inversion partner at different wavevectors. Considering a square lattice with only two identical atoms (A and B) in a unit cell with a small displacement between them along the $z$ direction [see Fig. 1(a)], such a buckling structure has a p4/nmm layer group with glide mirror reflection {$M_z$|(1/2,1/2,0)}, and screw axis operations {$C_{2x}$|(1/2,0,0)} and {$C_{2y}$|(0,1/2,0)}. In addition, the buckling also creates opposite local polar fields felt by each sector (here each atom forms a sector) along $\pm z$ direction, rendering a Rashba-type HSP system. We hereby construct a single-orbital (e.g., $s$, $p_z$ or $d_{z^2}$) tight-binding model (four bands) in the following (see Supplementary Section I for more details [23]),

$$H(\vec{k}) = t_1 \cos\frac{k_x}{2} \cos\frac{k_y}{2} \tau_x \sigma_0 + t_2 (\cos k_x + \cos k_y) \tau_0 \sigma_0$$
$$+ \lambda_R (\sin k_x \sigma_y - \sin k_y \sigma_x) \tau_z, \qquad (1)$$

where $\tau$ and $\sigma$ are Pauli matrices describing the sector and spin degrees of freedom, respectively. The first (second) term of Eq. (1) describes the nearest (next nearest) neighbor hopping, while the third term presents Rashba SOC caused by local polar fields. The band structures of Eq. (1) is shown in Fig. 1(b). In the absence of SOC ($\lambda_R = 0$), only the second term of Eq. (1) survives along the BZ boundary, leading to double degenerate (excluding spin), i.e., the "band sticking" effect due to nonsymmorphic symmetry [34]. When including SOC, the bands along the BZ boundary generally split into two doubly-degenerate bands, except at the high-symmetry TRIM points X and M. This is because at these points one can find a nonsymmorphic symmetry fulfills the anticommutation relationship with inversion operator, leading to an extra two-fold degeneracy between two pairs of Kramers degeneracy, i.e., four-fold degeneracy [35,36].

Nonsymmorphic symmetry has caught extensive attention recently due to such band degeneracy effect, leading to new types of quasiparticle such as Dirac node [36-39], nodal-chain [40] and nodal-surface [41], etc. However, its impact on the spin textures as well as Bloch states of different sublattices has not been substantially explored yet [42]. In order to quantify and then minimize the mixture of wavefunction between different sectors, we define a quantity named sector polarization $P_{sec} = (\rho_A -$ $\rho_B)/(\rho_A + \rho_B)$, where $\rho_A$ ($\rho_B$) is the module squared wavefunction projected onto A (B) sector. The sector polarization $P_{sec}$ at different wavevectors is shown in Fig. 1(c). The most important feature is that at the BZ boundary X-M, $P_{sec}$ is pinned to its maximum value ±1, indicating zero interaction between A and B sectors. In sharp contrast, the sector polarization at Γ point is exactly zero because the contribution of A and B sector to the wavefunction is equal. These observations can be explained by the model Hamiltonian, where the off-diagonal matrix elements are contributed solely by the first term of Eq. (1) containing $\tau_x$. At the BZ boundary, the first term vanishes, and the Hamiltonian is block diagonal and thus the direct sum of two subspaces spanned by the two sectors. Therefore, the two eigenstates are either located at A or B sector but not their superposition, leading to maximum sector polarization. On the other hand, when the wavevector moves to Γ, the first term with $\tau_x$ of Eq. (1) becomes more predominant, leading to descending $P_{sec}$ with the electron density finally distributed equally in A and B sectors at Γ.

The direct consequence of full sector polarization is that the HSP also reaches its maximum value at the BZ boundary as shown in the sector-projected spin texture in Fig. 1(d). Both inversion-asymmetric sectors retain large but opposite HSP, guaranteed by global inversion symmetry. In sharp contrast, around the Γ point the HSP localized at one sector is almost fully compensated by its inversion partner because the corresponding wavefunction contains substantial mixture of the two sublattices, leading to vanishing $P_{sec}$. Interestingly, we note that around the TRIM points, the projected spin textures are Rashba-type at the M point but Dresselhaus-type at the X point, validating that the symmetry requirement of Rashba effect always suggest an accompanying Dresselhaus effect [2]. Specifically, the low energy effective $k \cdot p$ Hamiltonians expanded by the SOC term of Eq. (1) takes the form of $(k_x\sigma_y - k_y\sigma_x)\tau_z$ at M, while $(k_x\sigma_y + k_y\sigma_x)\tau_z$ at X, indicating Rashba- and Dresselhaus-type spin patterns, respectively.

The sector polarization for different SOC strengths (quantified by $\lambda_R/t_1$) is also shown in Fig. 1(c). The main difference lies in the descending trends from M and X to Γ. Along these directions, the sector polarization $P_{sec}$ increases with $\lambda_R/t_1$, indicating that in Eq. (1) the SOC term containing $\tau_z$ enhances the polarization while the inter-sector coupling term containing $\tau_x$ suppresses the polarization. Consequently, it is desirable to have the protected HSP effect forming a large domain throughout the BZ by tuning the SOC strength relative to the inter-sector coupling.

**B. Symmetry consideration**

For the realization of realistic materials that manifest strong Rashba HSP, we next construct some symmetry requirements as design principles and scan the stable quasi-2D material databases, including Materialsweb [19] and AiiDA [20], to select ideal candidates with our target property. The reason we choose such layered materials is to build a straightforward connection with our tight-binding model. Since Rashba HSP requires a principle axis that is favored by a layered structure, our screening process can be easily generalized to 3D Rashba hidden spin system (such as LaOBiS$_2$ [2] and BaNiS$_2$ [18,43]). We choose layer group (80 in total) to classify the symmetry of quasi-2D materials, which greatly narrows the range of 3D space groups (230 in total) needed to be considered. The correspondence between all the layer groups and space groups is provided in the process of classification, as shown in Supplementary Section II [23].

The initial symmetry conditions are inversion symmetry and nonsymmorphic symmetry which have been shown to protect HSP at the BZ boundary in our model calculation. This search yields 17 layer groups from 4 point groups $C_{2h}$, $D_{2h}$, $C_{4h}$ and $D_{4h}$, and 225 material candidates from the chosen databases (see Table I). For $C_{4h}$ point group, there is only one layer group p4/n, which requires at least eight atoms to construct. On the other hand, we note that p2$_1$/m11, pmmn, and p4/nmm (belong to $C_{2h}$, $D_{2h}$, and $D_{4h}$, respectively) are the three most common layer groups that can be constructed by only two identical atoms, representing a total of 156 materials. These prototype structures of the four point groups and their model Hamiltonians are provided in Supplementary Section III [23].

Among the four prototype structures p2$_1$/m11 ($C_{2h}$) has the lowest symmetry. Due to the absence of screw axis along the $x$ direction, the 4-fold degeneracy of the X point lifts, and the HSP along M-X is not enforced to its maximum value, as shown in Fig. 2(a). Interestingly, the spin polarization now has sizable $s_z$ components peaked along

M-X, in contrast to the other three representative layer groups in which the spin orientation is totally in-plane. This is because the reduced symmetry of p2$_1$/m11 leads to an additional term $sink_y\sigma_z\tau_z$ in the model Hamiltonian [23]. Thus, as shown in Fig. 2(b), the total HSP around M is still significant but experience a visible out-of-plane canting from $k_x$ to $k_y$ direction (37° for the tight-binding parameters in Fig. 2).

**C. Materials realization**

We next apply density functional theory (DFT) calculations, with the presence of SOC, on the material candidates selected by the symmetry principles described above. The details of the computational methodology are provided in Supplementary Section IV [23]. Particularly, we find that a family of rare-earth compounds LnIO (Ln = Pr, Nd, Ho, Tm and Lu) manifests strong HSP throughout most of the BZ, in excellent agreement with the predictions based on our single-orbital model. As an example, LuIO crystallizes in a tetragonal nonmagnetic structure with a nonsymmorphic layer group p4/nmm (No. L64), containing 6 atoms per unit cell, as shown in Fig. 3(a). The oxygen plane containing inversion center separates the unit cell into two LuI sectors (A and B) that are connected with each other by the inversion symmetry. The two sectors feel opposite polar fields generated by their local environments that lack inversion symmetry, indicating the Rashba HSP. Fig. 3(b) shows the band structure of LuIO with SOC. As discussed above, each band is at least doubly degenerate, while the nonsymmorphic symmetry guarantees the four-fold band crossing at M point. The conduction band minimum is in the vicinity of M point, with the projected atomic orbitals dominated by $d_{z^2}$ character of Lu atoms, fitting our single-orbital model very well. Therefore, we would expect the lowest two doubly-degenerate conduction bands form a pair of Rashba bands with opposite helical spin texture.

The sector-projected spin polarization is calculated by projecting the wave functions $\psi_n(\boldsymbol{k})$ with plane-wave expansion on the orbital basis (spherical harmonics) of each atomic site and summing for a given sector A or B that contains a number of sites (here is Lu and I), as written in the following:

$$S_n^{A(B)}(\boldsymbol{k}) = \sum_{i\in A(B)} \sum_{l,m} \langle\psi_n(\boldsymbol{k})|(\sigma\otimes|l,m,i\rangle\langle l,m,i|)|\psi_n(\boldsymbol{k})\rangle, \qquad (2)$$

where $|l, m, i\rangle$ is the angular momentum eigenstate of *i*-th atomic site. As shown in Fig. 3(c), we find large and opposite spin polarizations of the two LuI sectors for the lowest two conduction bands. Remarkably, we observe a strong Rashba-type HSP pattern around M forming a clear domain that spans nearly 80% of the whole BZ. In addition, the spin texture around the X point is Dresselhaus-type, while around Γ there is a small region with vanishing HSP due to the strong interaction between the two sectors. As discussed above, the inter-sector hopping term totally vanishes only along the BZ boundary X-M, and the size of HSP domain in the whole BZ depends on the ratio between SOC strength and hopping parameter between two sectors $\lambda_R/t_1$, rendering ideal HSP effects in the materials with relatively strong SOC.

## III. Discussion

HSP protected by nonsymmorphic symmetry manifests two sets of spin-splitting bands that are degenerate in energy but localized at different inversion-asymmetric sectors. It would be desirable to exploit such property for future spintronic applications, such as modified spin field effect transistor model [7] and spin-dependent photogalvanic devices [44]. Furthermore, it provides a platform for tuning the interaction between sectors to realize exotic physics. For example, time-reversal-invariant topological superconductivity, with gapless Majorana modes at the edge states, is proposed to exist in an interacting bilayer Rashba system [9]. Such a band model is nothing but a HSP system with a hybridization gap at the Rashba band-crossing point. As shown above, the strong interaction at the Γ point requires a thick slab as a buffer layer between the two Rashba layers, posing challenges to sample growing and integration [45]. In comparison, HSP system offer us an alternative way to achieve the required band structure for experimental realization. As shown in Fig. 3(d), a tensile strain along [110] direction of LuIO breaks the screw axis symmetry and thus opens a hybridization gap Δ at the M point. The gap is 13 meV with a monoclinic distortion of the lattice γ = 89° and increases monotonically with larger distortion. Our approach to achieve interacting Rashba bilayer bands can be used to combine with other design principles such as odd-parity pairing symmetry for screening of topological superconductivity [21].

Finally, we note that for the energy bands with multi-orbital feature, the central physics, i.e., sector polarization and the resultant strong HSP protected by nonsymmorphic symmetry, still persists along the BZ boundary. On the other hand, the spin configuration for the other parts of the BZ could form a more complicated pattern. This is because orbitals with different azimuthal angular momentum couple different spin textures, while the total sector-projected spin texture is the superposition of the contributions from all considered orbitals [46]. To illustrate this, we further derive a four-orbital model Hamiltonian ($s$, $p_x$, $p_y$ and $p_z$ orbitals) containing 16 bands and make correspondence to the representative materials from DFT calculations. One of the main features induced by multi-orbital nature is the retention of HSP around the Γ point for the bands with considerable $p_x$ and $p_y$ components, as shown in Supplementary Section V [23]. Recently, a theoretical proposal by using a 1D chain model claimed that HSP vanishes in the vicinity of TRIM points due to the coupling between sectors [47]. On the contrary, our results point out that by simply adding either nonsymmorphic symmetry or multi-orbital feature [10], a HSP system would manifest nontrivial spin pattern for each inversion-asymmetric sector around TRIM points. Thus, it naturally closes such debate around the practical significance of HSP, invoking further motivations to look for materials with remarkable hidden spin textures and technologically-relevant properties. Since the probing beam of angular-resolved photoemission spectroscopy distinguishes different sectors by penetrating depth, the measurement of momentum-resolved HSP is highly accessible by counting the difference between the numbers of spin-up and spin-down photoelectrons. Hence, the experimental validation of our predictions on the physical effects and material candidates is strongly called for.

**Acknowledgements**

We thank Prof. Alex Zunger, Jun-Wei Luo, Xiuwen Zhang, Dr. Wen Huang, Quansheng Wu and Zhongjia Chen for helpful discussions. This work is supported by the NSFC under Grant No. 11874195.

Table I. List of layer groups with inversion symmetry and nonsymmorphic symmetry, and the corresponding quasi-2D materials. *Layer group pmam (L40) and pmma (L41) distinguish each other by the direction of the glide mirror, and these two layer groups correspond to the same 3D space group Pmma.

| Point group | Layer group | Space group | # of materials | Representative materials |
|---|---|---|---|---|
| $C_{2h}$ | L15 p$2_1$/m11 | 11 P$2_1$/m | 49 | AgBr |
| | L16 p2/b11 | 13 P2/c | 8 | Pr$_2$I$_5$ |
| | L17 p$2_1$/b11 | 14 P$2_1$/c | 31 | AgF$_2$ |
| $D_{2h}$ | L38 pmaa | 49 Pccm | 0 | - |
| | L39 pban | 50 Pban | 1 | BaP$_2$(HO)$_4$ |
| | L40 pmam* | 51 Pmma | 10 | CuHgSeBr |
| | L41 pmma* | | | |
| | L42 pman | 53 Pmna | 7 | P |
| | L43 pbaa | 54 Pcca | 0 | - |
| | L44 pbam | 55 Pbam | 2 | WBr$_2$ |
| | L45 pbma | 57 Pbcm | 1 | SiO$_2$ |
| | L46 pmmn | 59 Pmmn | 43 | ZrTe$_5$ |
| | L48 cmme | 67 Cmme | 3 | TlF |
| $C_{4h}$ | L52 p4/n | 85 P4/n | 4 | MoPO$_5$ |
| $D_{4h}$ | L62 p4/nbm | 125 P4/nbm | 0 | - |
| | L63 p4/mbm | 127 P4/mbm | 1 | MoBr$_2$ |
| | L64 p4/nmm | 129 P4/nmm | 64 | LuIO |

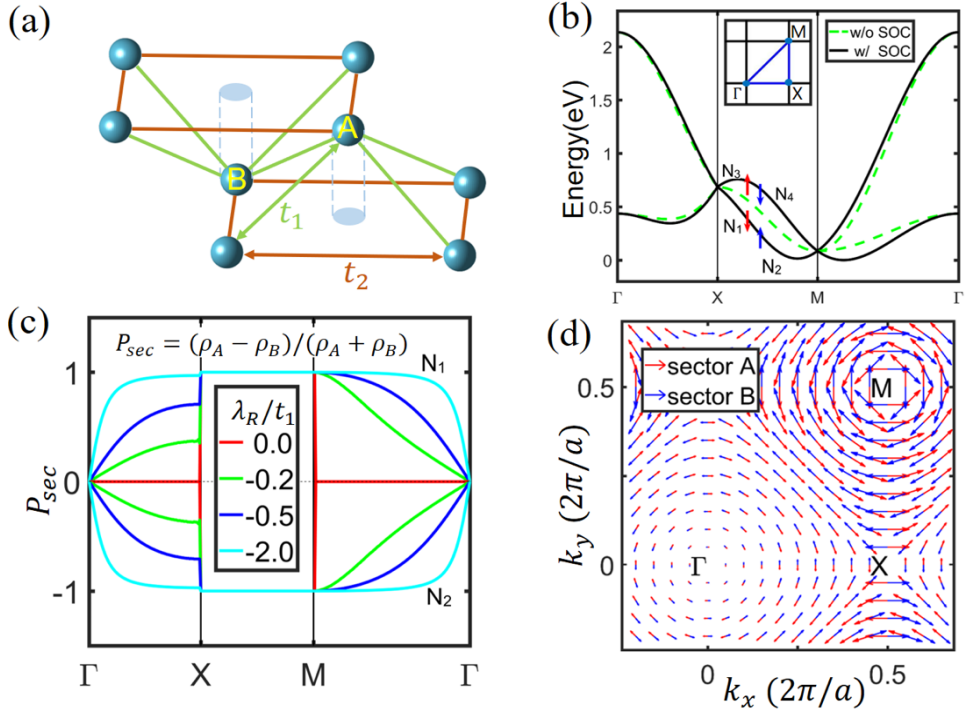

Fig. 1: (a) A square lattice with two identical atoms (A and B) in one unit cell. The nonsymmorphic symmetry is caused by the atomic displacement along the *z* direction. $t_1$ ($t_2$) is the nearest (next nearest) neighbor hopping parameter and $\lambda_R$ is the strength of Rashba SOC induced by the local polar field. (b) Band structure of this square lattice without and with SOC ($\lambda_R/t_1 = -0.2$). (c) Sector polarization ($P_{sec}$) of the lowest two doubly-degenerate bands $N_1$ and $N_2$ for different strength of SOC. To avoid the gauge problem, we applied a negligible electric field (i.e., an on-site potential difference of $10^{-8}$ eV between A and B atom) to break the band degeneracy. (d) Projected spin texture onto sector A (red) and B (blue) for the lowest two bands $N_1$ and $N_2$ ($\lambda_R/t_1 = -0.2$).

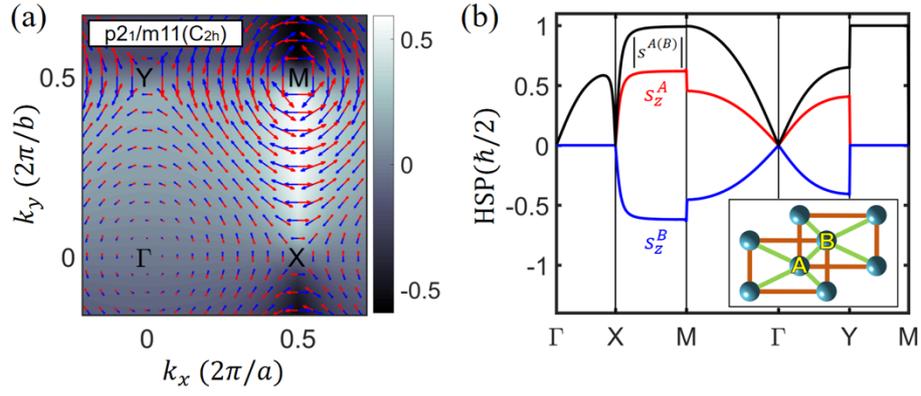

Fig. 2: (a) Projected spin texture onto sector A (red) and B (blue) for the lowest two bands ($\lambda_R/t_1 = -0.5$) of p2$_1$/m11 structure (C$_{2h}$). (b) Total HSP and the $s_z$ component of the two sectors. The crystal structure is shown in the inset of panel (b).

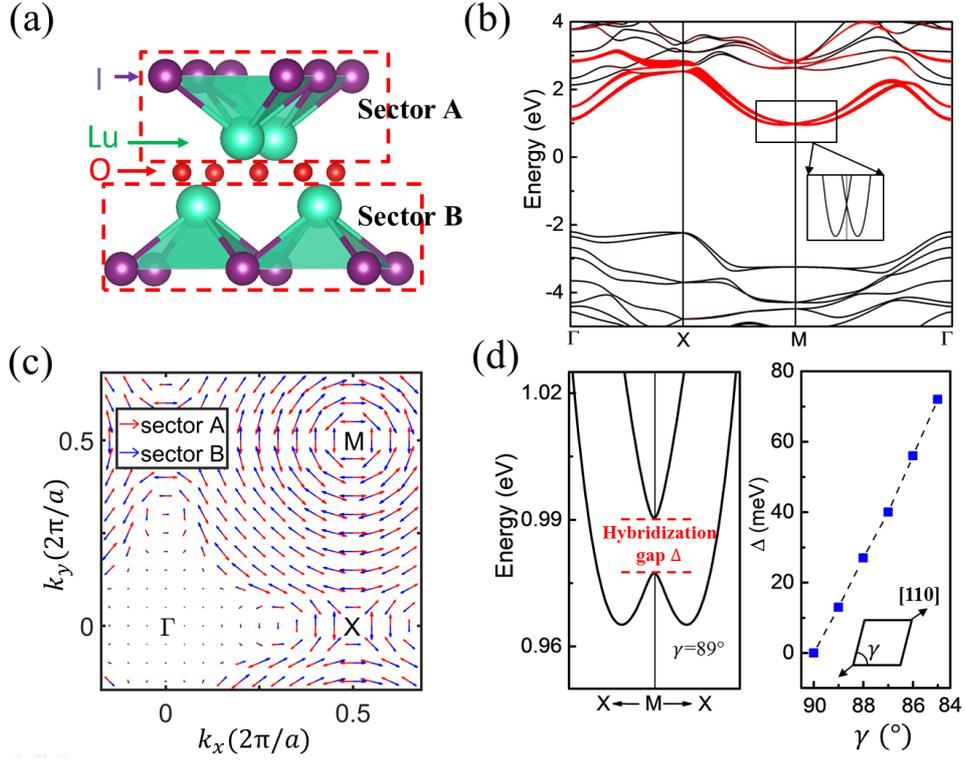

Fig. 3: (a) Crystal structure and two local sectors of LuIO. (b) Band structure (with SOC) with projection on to Lu-$d_{z2}$ orbital. (c) Projected spin texture onto sector A (red) and B (blue) for the lowest two conduction bands. (d) Hybridization gap at the M point induced by tensile strain along [110] direction (left), and its evolution along with the monoclinic distortion γ (right).